\documentclass[showpacs,preprintnumbers,amsmath,amssymb,aps,prl,preprint]{revtex4}
\usepackage{graphicx}
\usepackage{dcolumn}
\usepackage{bm}
\usepackage{amsmath}
\usepackage{color}

\begin{document}

\title{Theory of dissociative recombination of highly-symmetric polyatomic ions}
\author{Nicolas Douguet$^{1,2}$, Ann E. Orel$^2$, Chris H. Greene$^3$ and Viatcheslav Kokoouline$^{1}$}
\affiliation{$^1$Department of Physics, University of Central Florida, Orlando, FL 32816, USA\\
$^2$Department of CHMS, University of California at Davis, Davis, CA 95616, USA\\
$^3$Department of Physics and JILA, University of Colorado, Boulder, CO 80309, USA}

\begin{abstract} 
A general first-principles theory of  dissociative recombination is developed for highly-symmetric molecular ions and applied to H$_3$O$^{+}$ and CH$_3^+$, which play an important role in astrophysical, combustion, and laboratory plasma environments. The theoretical cross-sections obtained for the dissociative recombination of the two ions are in good agreement with existing experimental data from storage ring experiments.
\end{abstract}


\maketitle
The dissociative recombination (DR) of H$_3$O$^+$ and CH$_3^+$ ions with electrons is an important process in combustion, astrophysical, and laboratory plasma environment. For example, DR of H$_3$O$^+$ is believed to be the major source for water production in interstellar clouds and tails of comets, while the CH$_3^+$ ion triggers a series of reactions at the origin of formation of large hydrocarbons in the interstellar space. 

Description of DR of triatomic molecular ions with low-energy ($<$2 eV) electrons, developed in recent years \cite{kokoouline03b,kokoouline11a}, takes into account vibrational, rotational, and nuclear spin degrees of freedom. It provides detailed information about the recombination mechanism, about which rotational and vibrational states of the molecule participate, and about the final distribution of products. However, the full implementation of the method is very demanding numerically, if not impossible for four-atomic molecules such as H$_3$O$^+$ or CH$_3^+$, mainly due to the number of degrees of freedom involved. For H$_3^+$ \cite{greene06,jungen09,kokoouline11a} and HCO$^+$ \cite{mikhailov06,douguet08b,douguet09}, a simpler approach has been developed, in which one retains only such ingredients that have been found in the detailed quantum-mechanical study to represent the main DR mechanism. As a result, the simpler approach has proven to be successful and to agree well with the detailed theory and experiment. 

The present study takes another step in the development of the low-energy DR theory by generalizing it to treat other highly-symmetric ions, and by applying it to H$_3$O$^+$ and CH$_3^+$. The method implemented is based on ideas developed in earlier DR studies \cite{mikhailov06,douguet08b,jungen08a,jungen08b} but it is significantly generalized and accounts at present for vibronic coupling that has been neglected in the earlier studies. This coupling is small in H$_3^+$ but appears to be important in ions with many electrons. As in Refs. \cite{mikhailov06,douguet08b}, the starting point of the approach is the set of \textit{ab initio} potential surfaces of the neutral molecule and of the corresponding molecular ion, which are obtained numerically near the equilibrium position of the ion.

Our study of DR in H$_3^+$ showed that at temperatures $T>$100 K the overall resonance-averaged DR rate is mainly determined by the capture of the incident electron into Rydberg states attached to excited vibrational levels of the molecular ion. Inclusion of the rotational structure at modestly higher temperatures does not appreciably change the overall DR rate. Therefore, in the simplified method we neglect rotation and assume that the excited vibrational modes are responsible for the electron capture. For molecules heavier than H$_3^+$, the approach should be valid even for temperatures less than 100 K. Another assumption in this approach is that predissociation is faster than autoionization once the electron has been captured by the ion. This appears to be justified for the most of polyatomic ions \cite{mikhailov06,jungen10}. With these assumptions, the DR cross-section $\sigma(E_{el})$ for capture of the electron by an ion initially on its vibrational level $v'$ via the temporary capture into a Rydberg state associated to the vibrational ionic level $v$
is given by Eq. (14) of Ref. \cite{mikhailov06} 
\begin{equation}
\label{eq:cs0}
\langle\sigma(E_{el})\rangle=\frac{2a_0\pi^2}{e^2 k^2}
\Gamma_{v}
\nu^3\,,
\end{equation}
where $E_{el}=(\hbar k)^2/(2m_e)$, $k$, $e$, and $m_e$ are the kinetic energy, wave vector, charge, the mass of the incident electron, $a_0$ is the Bohr radius,  $\Gamma_{v}$ is the width of the autoionization resonance for the neutral molecule associated with the capture of the electron in the vibrational state $v$ of the ionic core, $\nu$ is the effective quantum number of the electronic wave function of the resonance. (Notice that symbols $v$ and $\nu$ are different. Also, an implicit assumption here is that a single electronic ionization threshold is relevant, whereby only one value of $\nu$ is relevant.) Brackets $\langle\rangle$ symbolically mean that the cross-section is averaged over the Rydberg series of resonances associated with the same vibrational state  $v$ of the ion \cite{mikhailov06} as well as over the rotational structure of the molecule. Below we omit the brackets for simplicity, but the  displayed cross-sections below are assumed to be averaged over the Rydberg series. If there are several vibrational states contributing to the cross-section for the capture, then a corresponding sum over $v$ should be evaluated.

The above formula allows a quick estimation of the DR cross-section, if experimental spectroscopic data (as in \cite{jungen09}) or {\it ab initio} data (as in Refs. \cite{kokoouline11a}) exists that can be used to determine $\Gamma_{v}$. As has been demonstrated in Refs. \cite{greene06,mikhailov06,jungen08b,jungen09,kokoouline11a}, this approach provides reasonable agreement with experiment for the molecular ions that have been considered to date. In order to calculate DR cross-sections for small and medium-size polyatomic ions with a modest computational {\it ab initio} effort, we rewrite the above formula in a different form.

With the above assumptions about averaging over rotational Rydberg series, the DR cross-section $\sigma(E_{el})$ for electron capture into vibrational level $v$ and electronic state $\Lambda$, with the ion initially in vibrational level $v'$ and the electron in state $\Lambda'$, can be written in terms of the scattering matrix $\langle v,\Lambda|\hat S|v',\Lambda'\rangle$ 
\begin{equation}
\label{eq:cs}
\sigma=\frac{\pi}{k^2}\vert\langle v,\Lambda|\hat S|v',\Lambda'\rangle\vert^2\,.
\end{equation}
The matrix element in the above equation can  be evaluated using the vibrational  frame transformation \cite{atabek74}
\begin{equation}
\langle v,\Lambda|\hat S|v',\Lambda'\rangle=\int d{\cal Q}\langle v|{\cal Q}\rangle S_{\Lambda,\Lambda'}({\cal Q})\langle{\cal Q} |v'\rangle\,.
\end{equation}
We assume that the ion is initially in the ground vibrational state $\vert v'\rangle= \vert 0\rangle$. We use the normal mode approximation for the vibration (as in \cite{jungen08b,jungen09})  with mode coordinates labeled as $q_i$ (${\cal Q}=\{q_1,q_2,\cdots\}$), the form $\hat S=\exp(2\pi i \hat\mu)$ of the scattering matrix, and with an expansion of $\mu_{\Lambda,\Lambda'}({\cal Q})$ in a Taylor series around the equilibrium configuration ${\cal Q}_0$  of the ion.  This yields
\begin{equation}
\label{eq:taylor}
\mu_{\Lambda,\Lambda'}({\cal Q})=\mu_{\Lambda,\Lambda'}({\cal Q}_0)+\sum_i \frac{\partial \mu_{\Lambda,\Lambda'}}{\partial q_i}q_i+\dots
\end{equation}
Here we use dimensionless normal modes $q_1, q_2,\cdots$, which are related to the length-unit normal coordinates $S_1, S_2,\cdots$ as $q_i=S_i\sqrt{\mu_{red}\omega/\hbar}$, where $\mu_{red}$ and $\omega$ are the reduced mass and the frequency of the normal mode. Retaining only zero- and first-order terms in the above expansion, Eq. (\ref{eq:cs}) takes the form \cite{kokoouline11a}
\begin{eqnarray}
\label{eq:cs2}
 \sigma_i(E_{el})=\frac{4\pi^3}{k^2}\left({\frac{\partial \mu_{\Lambda,\Lambda'}}{\partial q_i}}\right)^2\vert\langle v_i|\hat q_i|0\rangle\vert^2\,.
\end{eqnarray}
The index $i$ in $v_i$ and $\sigma_i$ is used to stress that the capture occurs into the $q_i$ vibrational mode excited by one vibrational quantum $v_i$. It is more convenient to use effective quantum numbers $\nu({\cal Q})=n-\mu({\cal Q})$, where $n$ is the principal quantum number, rather than quantum defects.  In the harmonic oscillator approximation, the matrix element $\langle v_i|\hat q_i|0\rangle=\delta_{v_i,1}/\sqrt{2}$. This gives
\begin{eqnarray}
\label{eq:cs3}
 \sigma_i(E_{el})=\frac{2\pi^3}{k^2}\left({\frac{\partial \nu_{\Lambda,\Lambda'}}{\partial q_i}}\right)^2 \theta(\hbar\omega_i-E_{el})g\delta_{v_i,1}\,.
\end{eqnarray}
The fact that the matrix element $\langle v_i|\hat q_i|0\rangle$ is nonzero only for $v_i=1$ provides a rationalization for the general observation that electron capture typically occurs into the lowest vibrational level. Contributions $\sigma_i(E_{el})$ from all of the normal modes $q_i$ must be added together to obtain the total cross-section. A particular contribution $\sigma_i(E_{el})$ should be set to 0 (at this crudest level of approximation) once the energy $E_{el}$ of the incident electron reaches the energy of one $v_i$ quantum.   This results in a sharp drop of the cross-section once the energy of the incident electron reaches the vibrational threshold -- an effect that has been observed experimentally for several triatomic molecules. In Eq.~(\ref{eq:cs3}) this is accounted for by the Heaviside step function $\theta$. We have also included the degeneracy factor $g$.

The actual DR cross-section is, in general, smaller than the cross-section for electron capture obtained in the way described above. Autoionization following the capture should decrease it to some degree. Once the electron is captured, the branching ratio between predissociation and autoionization probabilities determines the actual DR cross-section. The predissociation probability is determined by the strength of coupling between the state $|1_i,\Lambda\rangle$ formed in the collision with other states  $|v_i'',\Lambda''\rangle$.  On the other hand, the ``complex multichannel resonance phenomenon'' familiar from multichannel Rydberg spectroscopy (MQDT) \cite{jungen80,aymar96}, can lead to an enhancement in the electron capture probability. In the previous detailed study of DR in H$_3^+$ it was found that the electron capture enhancement due to complex multichannel resonances compensates for the probability of autoionization. In particular, it was found that the rotational-vibrational coupling helps to channel the energy of the captured electron towards higher-excited vibrational states of the ion. This leads to a decrease in the autoionization probability. In other words, even in H$_3^+$, the vibrational modes of the H$_3^+ + e^-$ system are coupled strongly enough to make the autoionization/dissociation branching ratio to be small. In larger polyatomic molecules, such intra-molecular vibrational energy transfer mediated by vibronic coupling should be even more efficient, although we have not made detailed calculations to confirm this in our applications to H$_3$O$^+$ and CH$_3^+$. We anticipate that the upper-limit estimate of the DR cross-section developed in the present study should be close to the actual value.

In order to use Eq. (\ref{eq:cs2}) we have determined the potential energy surfaces $U(\cal Q)$ of a few excited electronic states of the neutral molecule near the equilibrium position ${\cal Q}_0$ of the ion. Then, the effective quantum numbers $\nu(\cal Q)$ are obtained from the Rydberg formula $ U({\cal Q})=U^+({\cal Q})-{e^2}/(2a_0{\nu^2(\cal Q)})$, $U^+(\cal Q)$ being the electronic energy of the ion.

\begin{figure}
\includegraphics[width=14cm]{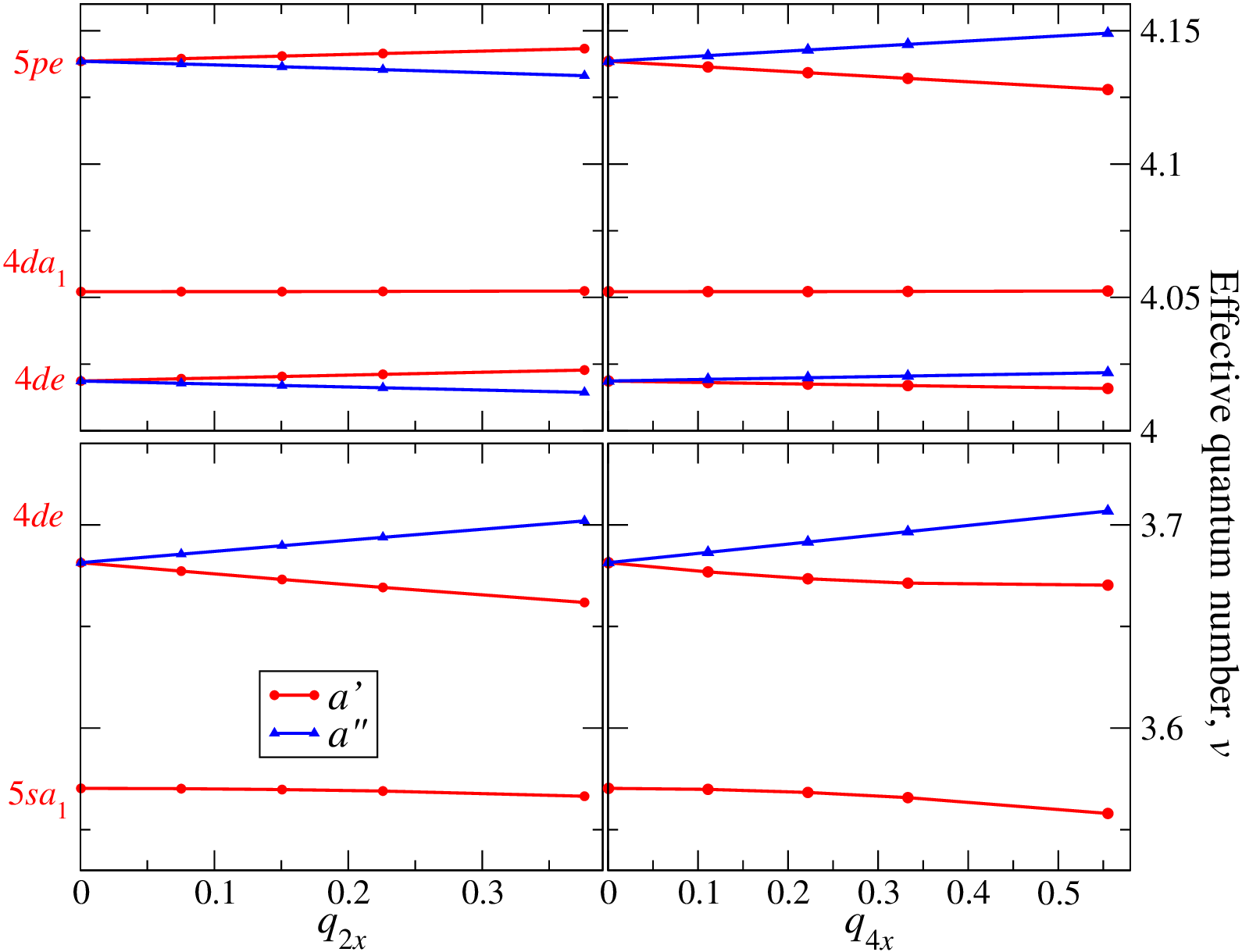}
\caption{H$_3$O effective quantum numbers $\nu({\cal Q})$. The two left panels show the dependence of $\nu$ for several electronic states of H$_3$O as a function of the $x$-component of the first doubly-degenerate vibrational mode $q_2$ of H$_3$O$^+$. The right panels show the same states for the second doubly-degenerate vibrational mode $q_4$. For $q_x\ne0$ electronic states are characterized by $C_s$ irreducible representations, i.e. $a'$ or $a''$.}
\label{fig:q2_q4_zoom}
\end{figure} 

The energies $U(\cal Q)$ and $U^+(\cal Q)$ and the related quantum defects and effective quantum numbers, which are obtained from {\it ab initio} calculations, correspond to the adiabatic representation where $\hat{\mu}$ and $\hat{\nu}$ matrices are diagonal but depend strongly on $\cal Q$, especially when $\cal Q$ is changing along one of the degenerate normal modes present in H$_3$O$^+$ and CH$_3^+$: Because such modes couple electronic states of the $e$ irreducible representation degenerate at ${\cal Q}_0$, the adiabatic approximation cannot be used in a straightforward manner. Fortunately, it is possible to construct a model Hamiltonian (similar to the one constructed in \cite{longuet61,koppel81,staib90}) for the electron-ion interaction in the diabatic representation. The same representation can then be used to construct the model matrices $\hat{\mu}(\cal Q)$ or $\hat{\nu}(\cal Q)$ , which are not diagonal (as in any diabatic representation) and contain a few parameters that can be fitted to reproduce {\it ab initio} energies.

The {\it ab initio} calculation has been made using the \textit{Columbus} software suite \cite{columbus}. In the calculation, the electronic wave function of the ionic core has been represented by the standard \texttt{cc-pvtz} \cite{dunning89} basis. For an accurate description of the excited electronic states of H$_3$O, \texttt{s,p}, and \texttt{d} diffuse universal Rydberg electronic orbitals optimized to represent the Rydberg wave functions \cite{kaufmann89} have been added to the ionic core orbitals. The number of additional Rydberg basis functions and the level of the configuration interaction was increased until the quantum defects converged to the needed accuracy. The origin $q_{2x}=q_{4x}=0$ corresponds to the equilibrium of H$_3$O$^+$.

The H$_3$O$^+$ ion at equilibrium has a $C_{3v}$ pyramidal form with three H atoms forming the equilateral base of the pyramid. It has four normal modes, and two of them are doubly degenerate, namely the second $q_2$ and the fourth $q_4$ modes (in order of increasing vibrational frequencies). We expect that the incident electron is captured by the ion due to the excitation of one of the degenerate modes, because the vibration along these modes breaks the $C_{3v}$ symmetry of the electronic states and, therefore, produces an additional vibronic coupling that is responsible for the large DR probability. Figure \ref{fig:q2_q4_zoom} shows the effective quantum numbers $\nu(q_{2x})$ and $\nu(q_{4x})$ on a limited interval of $\nu$ for several electronic states obtained from H$_3$O {\it ab initio} energies (calculations were performed up to $\nu=5$). 

%The figures shows only a limited interval of $\nu$; however, the present {\it ab initio} calculations are converged at least up to $\nu=5$.
%Changing the $x$-component of the degenerate modes keeps the symmetry of %the distorted molecule to be $C_s$, theres the $y$-component would break %this symmetry also. 

\begin{table}[h]
\begin{tabular}{l|ll|ll|l|ll|ll}
\hline
\multicolumn{5}{c|}{H$_3$O$^+$} &\multicolumn{5}{|c}{CH$_3^+$}  \\
\hline
\multicolumn{1}{c|}{} &\multicolumn{2}{c|}{$q_2$}&\multicolumn{2}{c|}{$q_4$} &\multicolumn{1}{|c|}{} &\multicolumn{2}{c|}{$q_2$}&\multicolumn{2}{c}{$q_4$} \\
\multicolumn{1}{c|}{} &\multicolumn{2}{c|}{$\omega=1705$}&\multicolumn{2}{c|}{$\omega=3705$} &\multicolumn{1}{|c|}{} &\multicolumn{2}{c|}{$\omega=1448$}&\multicolumn{2}{c}{$\omega=3262$} \\
\hline
$\nu_{eq.}$ & $\kappa_{ee}$ & $\kappa_{ea}$ & $\kappa_{ee}$ & $\kappa_{ea}$ &$\nu_{eq.}$ & $\kappa_{ee}$ & $\kappa_{ea}$& $\kappa_{ee}$ & $\kappa_{ea}$  \\
\hline
3.681  & 0.056 & 0.029 &0.046&0.047&4.65&0.083&0.062&0.041&0.045\\
4.018  & 0.012     &  small &0.006&small&4.29&0.020&small&0.004&small\\
4.139  & 0.013  & small &0.019&small&&&\\
\hline
\end{tabular}
%\vspace{0.3cm}
\caption{The fitted coupling parameters. Lines of numerical values correspond to the triads of electronic states included in the present treatment. The $e$ pairs of the triads are shown in Figs. \ref{fig:q2_q4_zoom} and \ref{fig:CH3}. Frequencies $\omega$ are in cm$^{-1}$ and calculated using the normal mode approximation and, therefore, somewhat different from the experimental values. In this study we enumerate vibrational modes in the order of increasing energy. Spectroscopic notations are different: The present $q_2$ and $q_4$ modes for both ions are modes 4 and 3 correspondingly.}
\label{table:fit}
\end{table}

\begin{figure}
\includegraphics[width=14cm]{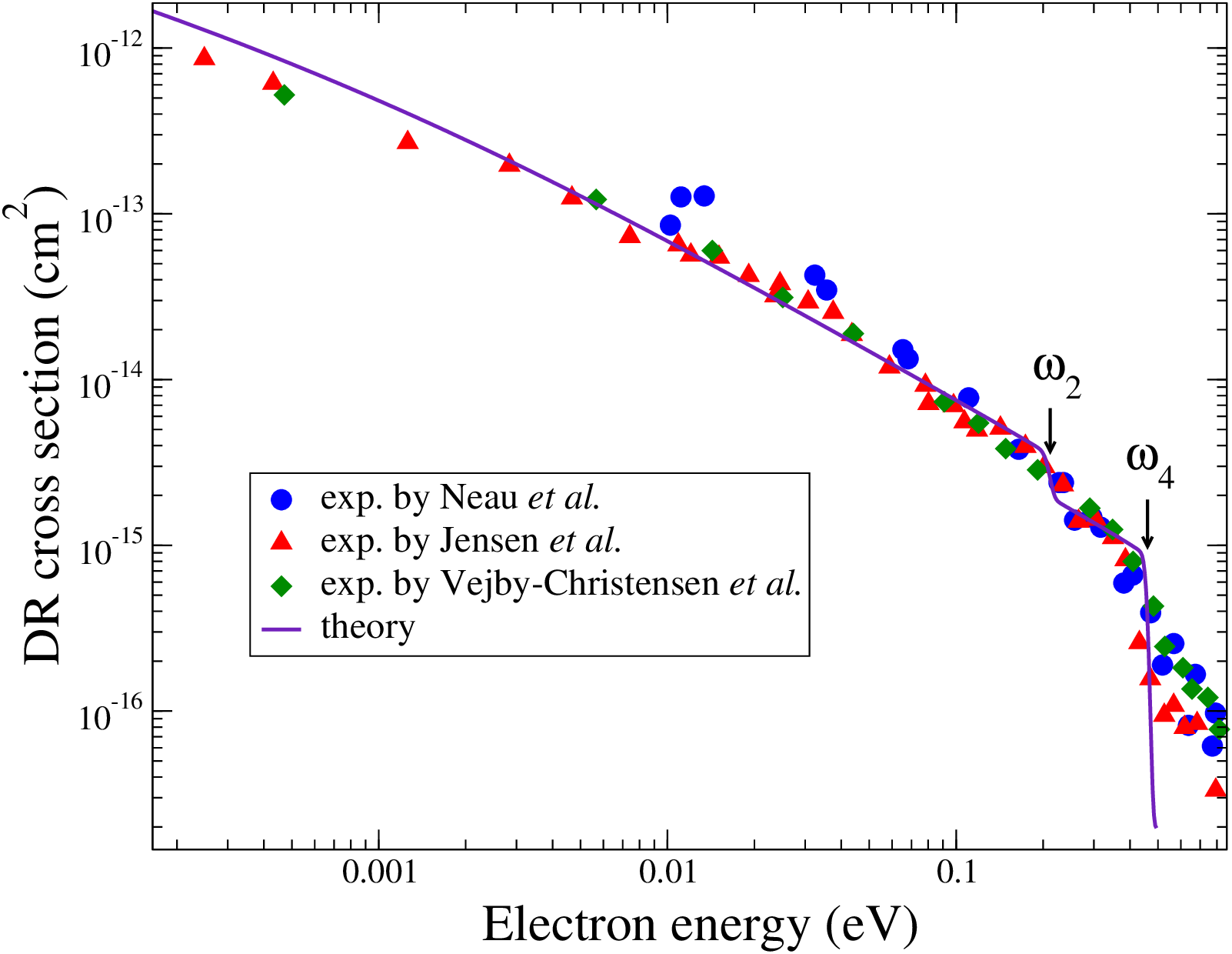}
\caption{Comparison of the present theoretical and experimental H$_3$O$^+$ DR cross-sections. The plotted experimental data are from Refs. \cite{christensen97,jensen00,neau00}. The theoretical curves include a convolution with an experimental anisotropic distribution $E_{||}=$0.1 meV and $E_{\perp}=2$ meV of electron energies. The theoretical curves also accounts for the toroidal correction as discussed in \cite{santos07}, which has a minor effect on the cross-section in the absence of autoionizing resonances. The arrows indicate threshold energies for the  $q_2$ and $q_4$ modes.}
\label{fig:cs}
\end{figure}

As one can see, three pairs of electronic states near $\nu({\cal Q}_0)=3.7,\ 4.02$, and $4.12$ are degenerate at the $C_{3v}$ geometry and are split by the Jahn-Teller coupling for distorted geometries. At ${\cal Q}_0$ these electronic states corresponds to $p\pi$, $d\pi$ and $d\delta$ states in the united atom approximation. Our initial intention was to obtain Jahn-Teller coupling parameters from the three pairs in order to obtain the DR cross-section as had been done in the previous study of H$_3^+$. However, as is clear from the right lower panel of the figure, the $4de$ state is also coupled to $5sa_1$. Therefore, we decided to include in the model Jahn-Teller Hamiltonian (and in the model matrix of effective quantum numbers), the corresponding coupling. The model matrix $\hat{\nu}$ in the diabatic basis of two $e$ and one $a_1$  electronic states can be written as a sum of unperturbed diagonal matrix $\hat{\nu}^{(0)}$, which is $q_2$- and $q_4$-independent, and the matrix of perturbation $\hat{\nu}^{(1)}$, which is linear in $q_2$ and $q_4$. The perturbation is caused by the $C_{3v}$ symmetry breaking. Diagonal elements of $\hat{\nu}^{(1)}$ are zero. Hence the total matrix $\hat\nu$ can be written as
\begin{equation}
\hat\nu(\rho,\varphi)=\left(
\begin{array}{ccc}
\nu^{(0)}_e &\kappa_{ea}\rho e^{i\varphi } &\kappa_{ee}\rho e^{-i\varphi } \\
\kappa_{ea}\rho e^{-i\varphi }& \nu^{(0)}_{a} & \kappa_{ea}\rho e^{i\varphi }\\
\kappa_{ee}\rho e^{i\varphi }& \kappa_{ea}\rho e^{-i\varphi }& \nu^{(0)}_e
\end{array}
\right) \,,
\label{eq:Hint_JT}
\end{equation}
where  $\rho$ and $\varphi$ are polar versions of the $x$- and $y$- components of $q_2$ or $q_4$. The parameters $\kappa_{ee}$ and $\kappa_{ea}$ are real, $\rho$- and $\varphi$-independent, and obtained by a fit to the {\it ab initio} results shown in Fig. \ref{fig:q2_q4_zoom}: The eigenvalues of the above matrix should be equal to the values obtained from {\it ab initio} energies. The fitted parameters $\kappa_{ee}$ and $\kappa_{ea}$ are given in Table \ref{table:fit} for both ions.

Using the fitted values $\kappa_{ee}$ and $\kappa_{ea}$, the contribution to the total DR cross-section from each triad of the $e$ and $a_1$ states for each vibrational mode is given by Eq. (\ref{eq:cs3}). The degeneracy factor has the value $g=2$ for the $e\sim e$ interaction: Only two projections (say the $p\pi^\pm$, $d\pi^\pm$, or $d\delta^\pm$ electronic states in the united-atom approximation) of the angular momentum of the incident electron on the molecular symmetry axis can excite vibration of the ionic core. For the $e\sim a$ interaction $g=4$: If the incident electron arrives in an $e$ electronic state included in the matrix of Eq. (\ref{eq:Hint_JT}), it can excite only one component of the doubly degenerate $q_2$ or $q_4$ modes by the $e\sim a$ coupling. There are two such electronic states $e$. However, if the electron arrives in the $a$ state included in the matrix of Eq. (\ref{eq:Hint_JT}), both vibrational components of the $q_2$ or $q_4$ modes can be excited. The $a$ state is coupled to the two $e$ electronic states and to both vibrational components of $q_2$ or $q_4$ with the value $g=4$.

\begin{figure}
\includegraphics[width=14cm]{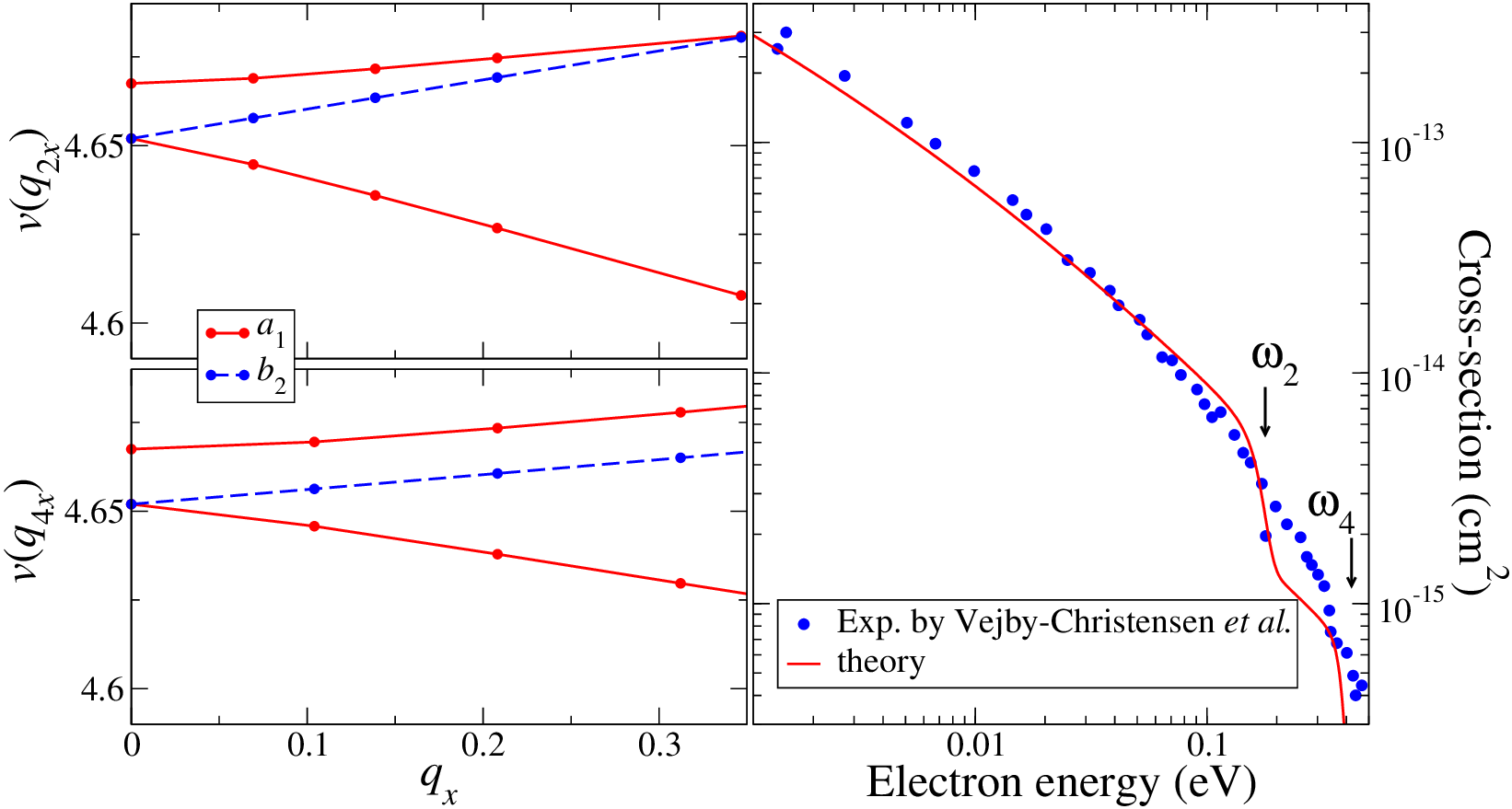}
\caption{ Effective quantum numbers and  DR cross-sections (theoretical and experimental \cite{christensen97}) for CH$_3^+$. Figure shows only a limited interval of $\nu(q_x)$ values to demonstrate that in the CH$_3^+$ case, the coupling between $e'$ and $a_1'$ ($D_{3h}$) electronic states is strong at distorted geometry because the quantum defects of the $e'$ and $a_1'$ electronic states at ${\cal Q}_0$ are almost degenerate. For $q_x\ne0$ the electronic states are characterized by the $C_{2v}$ irreducible representations, i.e. $a_1$, $a_2$, $b_1$, or $b_2$.}
\label{fig:CH3}
\end{figure} 

Figure \ref{fig:cs} shows the resulting DR cross-section for H$_3$O$^+$ and compares it with data from the ASTRID \cite{christensen97,jensen00} and CRYRING \cite{neau00} storage ring experiments. To compare with the experimental data, the non-uniform experimental distribution of collision energies between ions and electrons has been averaged over for the theoretical result.  We have also made a similar calculation for the CH$_3^+$ ion, which has $D_{3h}$ symmetry at the equilibrium and can be treated in the same way as H$_3$O$^+$: It has two $E'$  doubly degenerate modes that contribute to the DR cross-section. The result of the calculation is summarized in Fig. \ref{fig:CH3}. Here the theoretical cross-section also agrees well with the experimental data except the energy interval (0.2-0.4 eV) between the two vibrational thresholds. The disagreement could be explained by a presumably  high rotational temperature in the experiment. The rotational structure is not accounted for in the theoretical curve.

In conclusion, we stress that the present approach for theoretical determination of the DR cross-section in small polyatomic closed-shell ions at energies below 1 eV is general, simple in application, and gives good agreement with experiment as well as with the detailed theoretical study made for the benchmark problem of DR in H$_3^+$. 

We would like to thank Ioan Schneider for attracting our attention to study H$_3$O$^+$ and CH$_3^+$ and T. Oka for helpful comments on the manuscript. This work has been supported by the National Science Foundation and in part by the Department of Energy, Office of Science. VK also acknowledges support from the RTRA network {\it Triangle de la Physique}.

%\bibliography{../DR}

\end{document}